\documentclass[twocolumn,aps,prc,superscriptaddress,showpacs]{revtex4}

\usepackage{amssymb}
\usepackage{amsmath}
\usepackage{graphicx}
\usepackage[normalem]{ulem}
\usepackage{multirow}
\usepackage{appendix}
\usepackage{CJK}
\usepackage[usenames]{color}
\usepackage{bm}

\begin{document}
\begin{CJK*}{GBK}{song}

\title{
Revisit of directed flow in relativistic heavy-ion collisions from a multiphase transport model
}

\author{Chong-Qiang Guo}
\affiliation{Shanghai Institute of Applied Physics, Chinese Academy of Sciences, Shanghai 201800, China}
\affiliation{University of Chinese Academy of Sciences, Beijing 100049, China}
\author{Chun-Jian Zhang}
\affiliation{Shanghai Institute of Applied Physics, Chinese Academy of Sciences, Shanghai 201800, China}
\affiliation{University of Chinese Academy of Sciences, Beijing 100049, China}
\author{Jun Xu\footnote{Corresponding author: xujun@sinap.ac.cn}}
\affiliation{Shanghai Institute of Applied Physics, Chinese Academy of Sciences, Shanghai 201800, China}

\date{\today}

\begin{abstract}

We have revisited several interesting questions on how the
rapidity-odd directed flow is developed in relativistic
$^{197}$Au+$^{197}$Au collisions at $\sqrt{s_{NN}}$ = 200 and 39 GeV
based on a multiphase transport model. As the partonic phase evolves
with time, the slope of the parton directed flow at midrapidity
region changes from negative to positive as a result of the later
dynamics at 200 GeV, while it remains negative at 39 GeV due to the
shorter life time of the partonic phase. The directed flow splitting
for various quark species due to their different initial
eccentricities is observed at 39 GeV, while the splitting is very
small at 200 GeV. From a dynamical coalescence algorithm with Wigner
functions, we found that the directed flow of hadrons is a result of
competition between the coalescence in momentum and coordinate space
as well as further modifications by the hadronic rescatterings.

\end{abstract}

\pacs{25.75.-q, 
      25.75.Ld, 
      24.10.Lx  
}

\maketitle

\section{Introduction}

The main purpose of relativistic heavy-ion collision experiments is
to study the properties of the quark-gluon plasma
(QGP)~\cite{Ars05,Bac05,Ada05a,Adc05} and to understand the
hadron-quark phase transition. The anisotropic flow, defined as
$v_n=\langle \cos[n(\phi-\Psi_n)] \rangle$ with $\phi$ being the
particle azimuthal angle in momentum space with respect to the event
plane $\Psi_n$ and $\langle ... \rangle$ denoting the event average,
is an important observable in characterizing how the anisotropy in
the initial coordinate space develops into that in the final
momentum space, as a result of the strong interaction in the QGP
matter created in relativistic heavy-ion collisions. The first-order
anisotropic flow is named as the directed flow ($v_1$) (see
Ref.~\cite{v11} for a recent review), and it contains the
rapidity-odd component and the rapidity-even component. The
rapidity-odd component $v_1^{odd}(y) = -v_1^{odd}(-y)$, which is
traditionally called the sideward flow, is attributed to the
collective sidewards deflection of particles. The rapidity-even
component $v_1^{even}(y) = v_1^{even}(-y)$ was realized
recently~\cite{voe1,voe2}, and it is attributed to the
event-by-event fluctuation in the initial state of the colliding
nuclei. In the present study we only talk about the rapidity-odd
component of the directed flow.

Recently, RHIC-STAR Collaboration have reported the directed flow of
protons and pions in the beam-energy-scan program~\cite{RHIC}. It
has been found that the slope of the net-proton directed flow
changes sign twice between $\sqrt{s_{NN}}$ = 11.5 GeV and 39 GeV,
and has a minimum between $\sqrt{s_{NN}}$ = 11.5 GeV and 19.6 GeV.
Besides, splittings of the directed flow between protons and
antiprotons as well as that between $\pi^+$ and $\pi^-$ were
observed at lower collision energies but become small at higher
collision energies. Efforts have been made in understanding the
above directed flow data~\cite{Ste14,Kon14,Iva15,Bat16,Nar16,Nar17}.
In this study we investigate several interesting topics relevant to
the directed flow in relativistic heavy-ion collisions within a
multiphase transport (AMPT) model~\cite{AMPT}. Different from the
previous study~\cite{chenjy}, we have studied the non-monotonic
evolution of the directed flow in the partonic phase, the splitting
of the directed flow between different quark species, and the
effects of the hadronization and hadronic evolution on the directed
flow. The study helps clarify how the directed flow is developed or
modified at different stages in relativistic heavy-ion collisions,
and is useful in understanding the directed flows at different
collision energies. The rest of the paper is organized as follows.
Section~\ref{sec:2} provides a brief introduction of the AMPT model.
The detailed analysis and discussions of the directed flow results
are given in Sec.~\ref{sec:3}. Finally, a summary and outlook is
given in Sec.~\ref{sec:4}.

\section{\label{sec:2} THE AMPT MODEL}

The string melting version of the AMPT model~\cite{AMPT}, which is
used in the present study, mainly consists of four parts: the
initial condition generated by Heavy Ion Jet Interaction Generator
(HIJING) model \cite{hijing}, the partonic evolution described by
Zhang's parton cascade (ZPC) model \cite{zpc}, a coalescence model
to describe the hadronization process, and the hadronic evolution
described by a relativistic transport (ART) model \cite{art}. The
HIJING model generates hadrons with proton-proton scatterings as the
building brick together with the nuclear shadowing effect and the
Glauber geometry for the colliding nuclei at relativistic energies.
The initial phase-space distribution of partons is generated by
melting hadrons produced by elastic and inelastic scatterings of participant nucleons in HIJING. The partonic interaction in the ZPC model
is described by the partonic two-body elastic scatterings with the
differential cross section given by
\begin{equation}\label{sigma}
\frac{d\sigma}{dt} \approx \frac{9\pi{\alpha_s^2}}{2(t-\mu^2)^2},
\end{equation}
where $t$ is the standard Mandelstam variable for four-momentum
transfer. In the present study we set the strong coupling constant
$\alpha_s$ to be 0.47 and the parton screening mass $\mu$ to be
3.2264 fm$^{-1}$, leading to the total cross section of 3 mb. Partons
freeze out continuously after their last scatterings, and the
hadronization is treated according to the freeze-out phase-space
distribution of all partons. The hadronization in AMPT is described
by a spatial coalescence model which allows a pair of nearest quark
and antiquark to form a meson and three nearest quarks (antiquarks)
to form a baryon (antibaryon), with the mass and species of the
hadron determined by the invariant mass and the flavors of these
constituent partons. In the present study we do the coalescence for baryons and antibaryons before that for mesons. In this way there are more combinations of quarks (antiquarks) close in phase space to form baryons (antibaryons), which helps to give a smooth $v_1$, while meson $v_1$ is not much affected since there are still plenty of choices for daughter quarks/antiquarks to form mesons. In order to see the effect of a more realistic
coalescence on the directed flow, we have also checked with the
dynamical coalescence \cite{dynamic} based on the Wigner function
calculation detailed in Sec.~\ref{sec:3}B and \ref{app}.
The spatial coalescence in the AMPT model is followed by the ART
model that contains various elastic, inelastic, and decay channels
to describe the hadronic evolution.

\section{\label{sec:3} ANALYSIS AND RESULTS}

In the present study, we employ the AMPT model to investigate the
directed flow in midcentral ($\rm{b} = 5$ fm) $^{197}$Au+$^{197}$Au
collisions at $\sqrt{s_{NN}}$ = 200 and 39 GeV, corresponding to the
top RHIC energy and a typical energy in the beam-energy-scan
program. Typically, we focus on the time evolution of $v_1$, the
splitting of $v_1$ for various particle species, and the
hadronization effect on $v_1$. The directed flow is calculated from
$v_1=\langle \cos(\phi-\Psi_{RP}) \rangle$ with respect to the
theoretical reaction plane $\Psi_{RP}=0$.

\subsection{Time evolution of $v_1$ in the partonic phase}

\begin{figure}[h]
\includegraphics[scale=0.35]{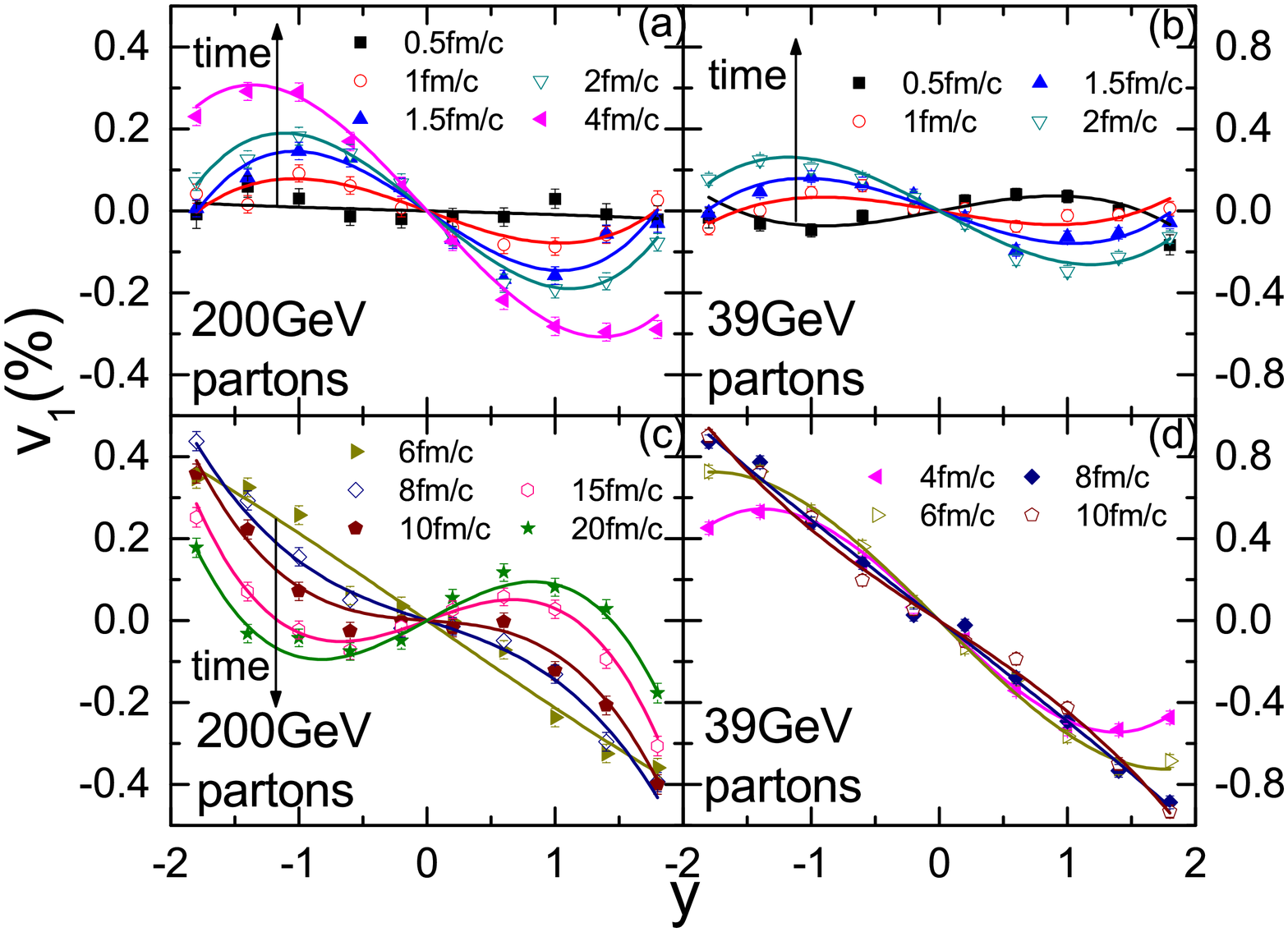}
\caption{(Color online) Directed flow ($v_1$) of partons versus
rapidity ($y$) at different time steps in midcentral Au+Au
collisions at $\sqrt{s_{NN}}$ = 200 GeV (left) and 39 GeV (right). The upper panels show the behavior in early stages, while the lower panels show that in later stages.}
\label{F3}
\end{figure}

The directed flows of partons in midcentral Au+Au collisions at
$\sqrt{s_{NN}}$ = 200 GeV and 39 GeV at different time steps are
displayed in Fig.~\ref{F3}, where the upper (lower) panels show the evolutions in early (later) stages, with the solid lines from a cubic fit of
$v_1(y) = F_1 y + F_3y^3$. The initial $v_1$ at both collision
energies are very small as expected. It is seen that the
slope of the directed flow at $\sqrt{s_{NN}}$ = 200 GeV grows to a
maximum negative value in early stages ($t<4$ fm/c), and then gradually becomes
positive in later stages, while that at $\sqrt{s_{NN}}$ = 39 GeV grows to a maximum
negative value and becomes saturated. The maximum
slope is larger at 39 GeV than at 200 GeV. By monitoring the density
evolution, we found that the strong scatterings among partons mostly
end around $4 \sim 6$ fm/c. However, it is seen that the later
dynamics reverses the slope of the directed flow at $\sqrt{s_{NN}}$
= 200 GeV but is unable to reverse that at $\sqrt{s_{NN}}$ = 39 GeV,
due to the shorter life time of the partonic phase at lower
collision energies. The non-monotonic behavior of the directed flow
was also observed in Refs.~\cite{Liu99,Ste14}. The feature mentioned
above is qualitatively consistent with the less negative $v_1$ slope
at midrapidities at higher collision energies observed by
PHOBOS~\cite{PHOBOSv1} and STAR~\cite{STARv1,RHIC} Collaborations.

\begin{figure}[h]
\includegraphics[scale=0.32]{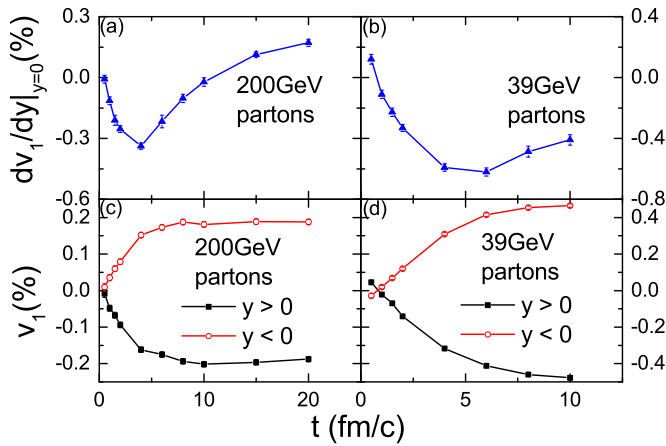}
\caption{(Color online) Time evolution of the parton directed flow slope ($dv_1/dy|_{y=0}$) (upper panels), and the integrated directed flow
($v_1$) at forward and backward rapidities (lower panels) in midcentral Au+Au
collisions at $\sqrt{s_{NN}}$ = 200 GeV [(a) and (c)] and 39 GeV [(b) and (d)].}
\label{F5}
\end{figure}

The time evolution of the directed flow slope at midrapidities is displayed in the upper panels of Fig.~\ref{F5}. It is clearly seen that at $\sqrt{s_{NN}}$ = 200 GeV the directed flow slope first drops to a negative value lower than $-0.3\%$ and than increases to a positive value of about $0.15\%$. At $\sqrt{s_{NN}}$ = 39 GeV, however, the directed flow slope drops to about $-0.6\%$ and the later dynamics only slightly modifies the slope. We have further displayed the integrated directed flow at forward and
backward rapidities as a function of time in the lower panels of Fig.~\ref{F5}. It is
interesting to see that the integrated $v_1$ at the forward
(backward) rapidity monotonically becomes more negative (positive)
as the system evolves, although the directed flow at different
rapidity regions changes in a complicated manner as shown in
Fig.~\ref{F3}. At $\sqrt{s_{NN}}$ = 200 GeV the integrated directed
flow becomes saturated at about 8 fm/c, while at $\sqrt{s_{NN}}$ =
39 GeV it is saturated at a later time. In addition, the magnitude
of the integrated $v_1$ is larger at lower collision energies.

We have further investigated the time evolution of the $v_1$ slope given the saturated integrated directed flow at both
forward and backward rapidities as shown in Fig.~\ref{F5}. We found
that the time evolution of $v_1$ is due to the transfer of
particles, which contribute positively or negatively to $v_1$, among
different rapidity regions. At the later stage of the partonic phase
at $\sqrt{s_{NN}}$ = 200 GeV, more particles that contribute to the
positive flow stay in the midrapidity region, while those contribute
to the negative flow move to larger rapidities. At $\sqrt{s_{NN}}$ =
39 GeV, the saturation of $v_1$ takes longer time while the life
time of the partonic phase is too short to reverse $v_1$, leading to
a negative slope at the freeze-out stage.

\subsection{Splitting of $v_1$ for various quark species}

The results discussed in the previous subsection are averaged over
all quark species. On the other hand, it is always observed that
there are splittings of quantities between particles and their
antiparticles as well as those between particles of different
isospin states, especially at lower collision energies. The typical
examples are splittings of the elliptic flow~\cite{RHICv2} and the
directed flow~\cite{RHIC} between protons and antiprotons as well as
those between $\pi^+$ and $\pi^-$.

\begin{figure}[h]
\includegraphics[scale=0.22]{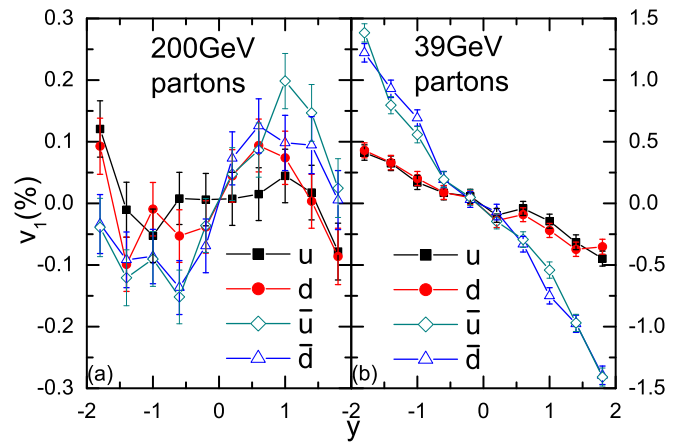}
\caption{(Color online) Final directed flow ($v_1$) of $u$, $d$, $\bar{u}$, and
$\bar{d}$ quarks versus rapidity ($y$) in midcentral Au+Au
collisions at $\sqrt{s_{NN}}$ = 200 GeV (a) and 39 GeV (b).}
\label{F7}
\end{figure}

Figure~\ref{F7} displays the directed flow of $u$ and $d$ quarks as well as their antiquarks at their freeze-out stage in midcentral Au+Au
collisions at $\sqrt{s_{NN}}$ = 200 GeV and 39 GeV. At
$\sqrt{s_{NN}}$ = 200 GeV the $v_1$ splitting between quarks and
antiquarks as well as that between $u$ and $d$ quarks are already
seen but the difference is comparable to the statistical error,
while at $\sqrt{s_{NN}}$ = 39 GeV it is clearly seen that
antiquarks have a more negative directed flow slope than quarks. We note that here quarks include produced and transported ones from initial inelastic and elastic scatterings of participant nucleons, respectively, while antiquarks are all produced from inelastic nucleon-nucleon scatterings. As shown in Refs.~\cite{Dun11,Guo12}, produced and transported particles generally have different collective flows. The splitting between directed flows of various quark species could
be partially responsible for the $v_1$ splitting between protons and
antiprotons as well as that between $\pi^+$ and $\pi^-$, as reported
in Ref.~\cite{RHIC}.

\begin{figure}[h]
\includegraphics[scale=0.22]{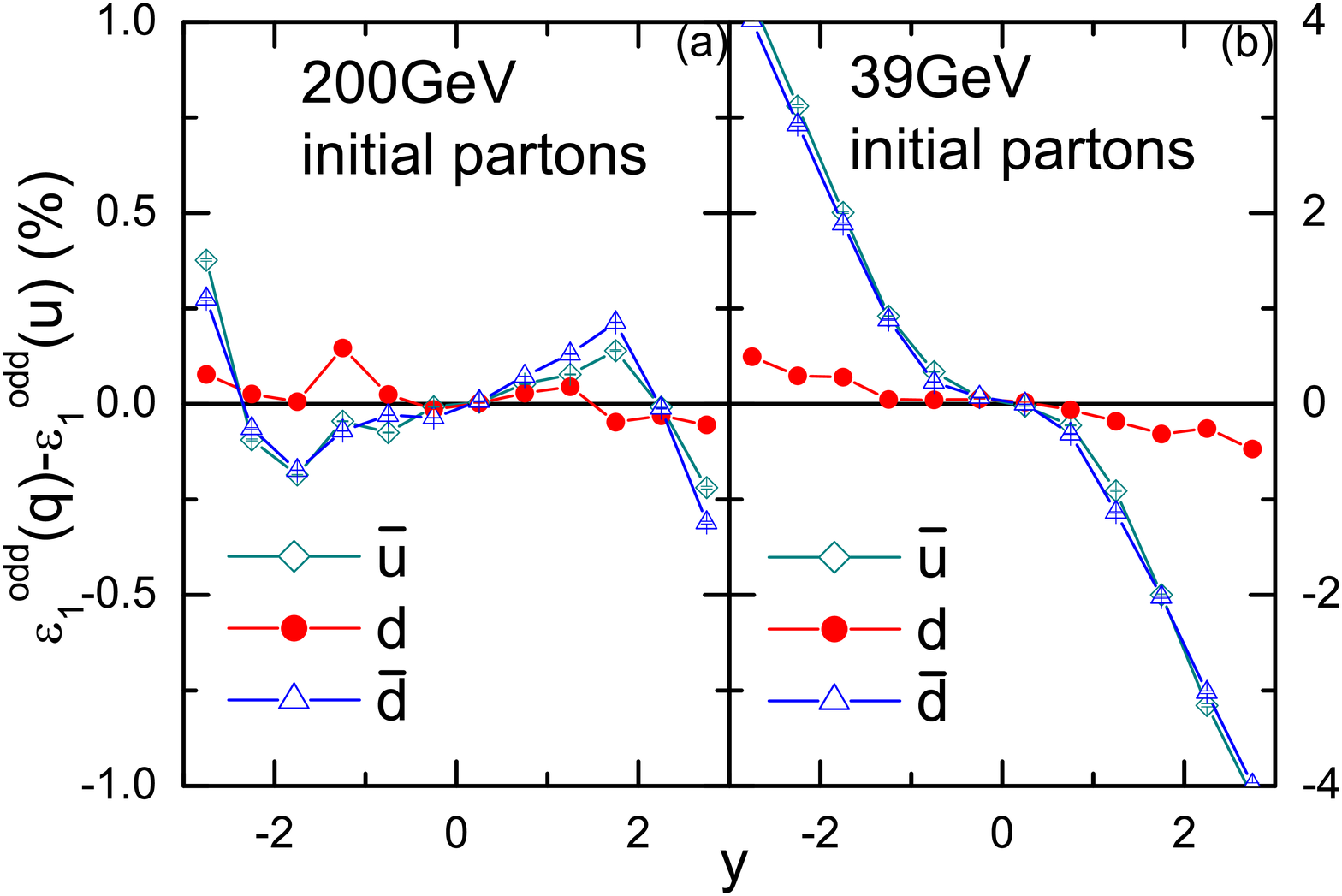}
\caption{(Color online) Initial first-order eccentricity
$\epsilon_1^{odd}$ of $d$, $\bar{u}$, and $\bar{d}$ quarks with respect to that
of $u$ quarks versus rapidity $y$ in midcentral Au+Au collisions at
$\sqrt{s_{NN}}$ = 200 GeV (a) and 39 GeV (b).} \label{F8}
\end{figure}

Since the dynamics in the partonic phase dominated by the parton
scattering cross section [Eq.~(\ref{sigma})] is independent of the
quark species, the splitting of the directed flows of various quark
species shown in Fig.~\ref{F7} can only be due to their different
initial eccentricities. The mechanism how the initial eccentricity
($\epsilon_n$) develops into the final elliptic flow ($v_2$) and
triangular flow ($v_3$) has been extensively studied (see, e.g.,
Refs.~\cite{epsilon1,epsilon2,epsilon3}). The response of the
rapidity-even directed flow to the rapidity-even $\epsilon_1^{even}$
was discussed in Refs.~\cite{voe1,voe2}. The initial rapidity-odd
$\epsilon_1^{odd}$ according to the particle azimuthal angle
$\phi_s$ in coordinate space can be calculated
as~\cite{Liu99,wiggle}
\begin{equation}\label{epsilon}
\epsilon_1^{odd}(y) = \langle \cos(\phi_s-\Psi_{RP}^{pp}) \rangle_y,
\end{equation}
with $\langle ... \rangle_y$ denoting the event average at a given
rapidity $y$, and we used the theoretical reaction plane
($\Psi_{RP}^{pp}=0$) consistent with the calculation of the
rapidity-odd directed flow. Figure \ref{F8} shows the rapidity
distribution of $\epsilon_1^{odd}(y)$ of $d$, $\bar{u}$, and $\bar{d}$ quarks
with respect to that of $u$ quarks at both $\sqrt{s_{NN}}$ = 200 GeV
and 39 GeV. It is seen that at $\sqrt{s_{NN}}$ = 200 GeV the initial
$\epsilon_1^{odd}$ has small difference especially between quarks and antiquarks, while at
$\sqrt{s_{NN}}$ = 39 GeV the difference in the slope with respect to
rapidity for different quark species is much larger. Since these
partons are melted from hadrons produced in HIJING, the difference is
attributed to the different production mechanisms of hadrons that
have different baryon or isospin charges. The different $v_1$ of $d$, $\bar{u}$,
and $\bar{d}$ quarks compared with that of $u$ quarks is attributed
to their different $\epsilon_1^{odd}$ at $\sqrt{s_{NN}}$ = 39 GeV.
In addition, we found the initial averaged $\epsilon_1^{odd}$ is slightly
larger at lower collision energies, responsible for the larger
saturated $v_1$ observed in Fig.~\ref{F5}.

\subsection{Effect of hadronization on $v_1$}

The directed flow of freeze-out partons discussed in the previous
subsections will be modified in the hadronization process. In the
present study we investigate the hadronization from a dynamical
coalescence model~\cite{dynamic,splitting} and the default spatial
coalescence model as in AMPT. In the dynamical coalescence model,
partons that are close in phase space have a larger probability to
form hadrons, while in the default spatial coalescence model in
AMPT, hadrons are formed by nearest combinations of partons in
coordinate space as discussed in Sec.~\ref{sec:2}, and all partons
are forced to be used up after hadronization.

In the dynamical coalescence model, the probability to form a hadron
is proportional to the parton Wigner function of that hadron, and the
proportional coefficient is the statistical factor by considering
the spin, flavor, and color degeneracies. For detailed formulaes we
refer the reader to \ref{app}. There are also other hadronization
mechanisms, such as the fragmentation of high-momentum partons in
transport models and the Cooper-Frye hadronization in hydrodynamic
models. However, the dynamical coalescence has already included the
main feature of quark recombination, which explains the number of
constituent quark scaling of collective flows~\cite{dynamic,naive1}
as one of the evidences of the formation of QGP. To speed up the calculation, we skip the parton combinations with large relative momenta, since their probabilities to form hadrons are small.

\begin{figure}[h]
\includegraphics[scale=0.38]{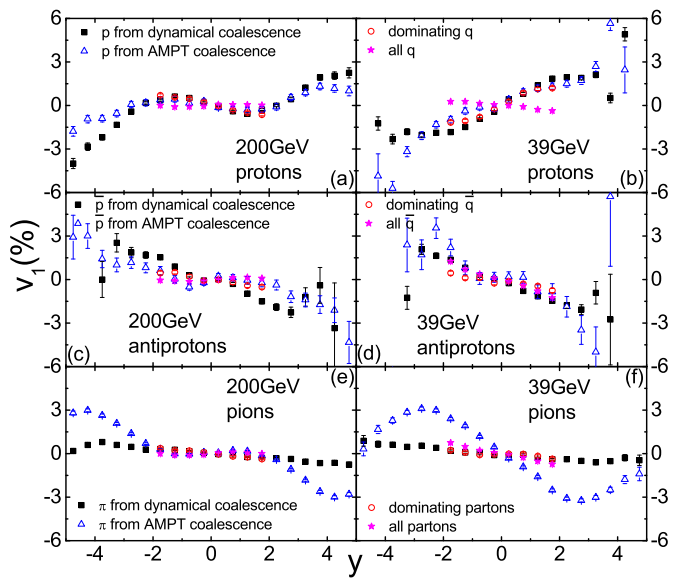}
\caption{(Color online) Directed flows ($v_1$) of protons (top), antiprotons (middle), and charged
pions (bottom) from the dynamical coalescence (squares) and from the default spatial
coalescence in the AMPT model (triangles), as well as $v_1$ of their dominating constituent
quarks or antiquarks weighted by the Wigner function (circles), and $v_1$ of all quarks or antiquarks (stars) at
their freeze-out stage as a function of rapidity in midcentral
Au+Au collisions at $\sqrt{s_{NN}}$ = 200 GeV (left) and 39
GeV (right). See text for details.} \label{F9}
\end{figure}

Figure~\ref{F9} displays the directed flows of protons (top panels), antiprotons (middle panels), and charged
pions (bottom panels) from the dynamical coalescence and from the default spatial
coalescence in AMPT, as well as that of their dominating constituent quarks or antiquarks
weighted by the Wigner function, and that of all quarks or antiquarks at their
freeze-out stage. The difference between $v_1$ from the dynamical
coalescence and that from the default spatial coalescence in AMPT is
observed, consistent with the statement in Ref.~\cite{Ste14} that
the directed flow is sensitive to the hadronization treatment. It is
interesting to see that the directed flow of constituent quarks (antiquarks)
weighted by the Wigner function is quite different from that of all
quarks (antiquarks) at their freeze-out stage. This means that in the dynamical
coalescence scenario only part of partons close in phase space
dominate the contribution of forming hadrons, while their directed
flow is quite different from the rest partons. On the other hand, the
directed flow slope of formed protons, antiprotons, or pions is different from that of their
constituent partons or the overall partons.
The difference is larger for baryons or antibaryons than for mesons. It is seen that the $v_1$
slope of protons from the dynamical coalescence changes its sign
compared with that of the overall quarks.

In order to understand how the coalescence modifies the directed
flow, we employ the pure momentum coalescence and pure coordinate
coalescence, by considering only the momentum or the coordinate part
in Eqs.~(\ref{fm}) and (\ref{fb}). This is similar to the limit of
choosing an infinitely large or a zero Gaussian width for the Wigner
function, respectively. With the pure momentum coalescence, we found
that the slope sign of $v_1$ near the midrapidity region is not
changed after coalescence, comparing with the parton directed flow
at freeze-out, consistent with the picture of the naive coalescence
scenario~\cite{naive1,naive2}, which leads to the number of constituent
quark scaling relation, i.e., $V_{1}(\bm{p}) \approx
2v_{1}({\bm{p}}/{2})$ for mesons and $\tilde{V}_{1}(\bm{p}) \approx
3v_{1}({\bm{p}}/{3})$ for baryons as detailed in \ref{naive}. With
the pure coordinate coalescence, the parton density distribution
becomes important, and the slope sign of $v_1$ is generally changed.
The hadron directed flow is a result of competition between the
coalescence in momentum and coordinate space, with the weight
determined by the Gaussian width fitted by the root-mean-square
radius of the hadron.

\begin{figure}[h]
\includegraphics[scale=0.22]{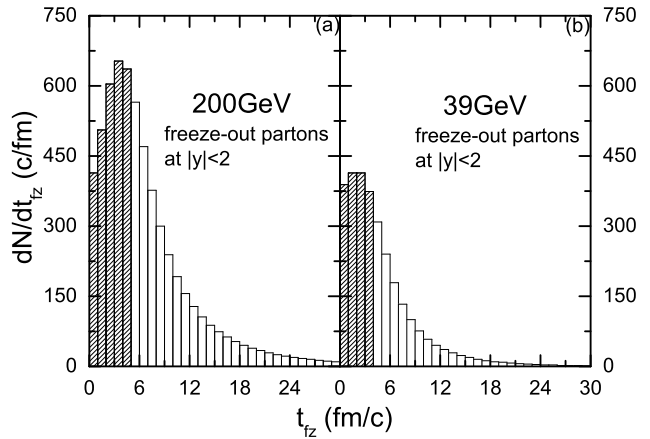}
\caption{Distribution of parton freeze-out time in midcentral Au +
Au collisions at $\sqrt{s_{NN}}$ = 200 GeV (a) and $\sqrt{s_{NN}}$ =
39 GeV (b), with the shadow indicating the early freeze-out partons
that dominate the hadron formation.} \label{F11}
\end{figure}

\begin{figure}[h]
\includegraphics[scale=0.2]{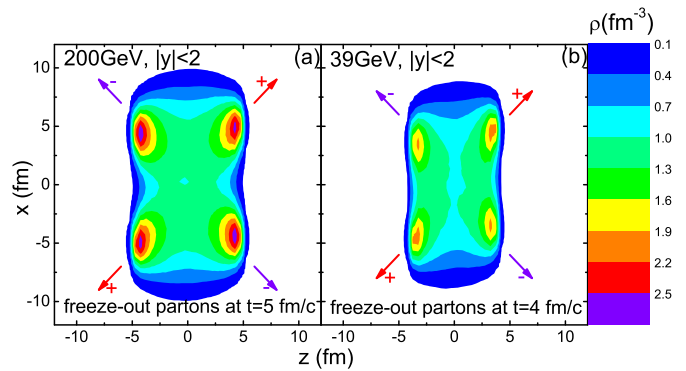}
\caption{(Color online) Density contours of freeze-out partons in
the midrapidity region in the x-o-z plane in midcentral Au + Au
collisions at $t=5$ fm/c for $\sqrt{s_{NN}}=$ 200 GeV  (a) and at
$t=4$ fm/c for $\sqrt{s_{NN}}=39$ GeV (b), with different colors
indicating the parton number densities. The particles that move in
the ``$+(-)$" direction contribute to the positive flow (antiflow).}
\label{F12}

\end{figure}

Here we further illustrate why the pure coordinate coalescence
generally changes the slope sign of the directed flow.
Figure~\ref{F11} displays the distribution of the parton freeze-out
time. In the dynamical coalescence scenario partons that freeze out
at a similar time are more likely to coalesce with each other. We
found that the early freeze-out partons, i.e., those freeze out
before $t_{fz}=5$ fm/c at 200 GeV and $t_{fz}=4$ fm/c at 39 GeV,
dominate the hadron formation, and their freeze-out times are
indicated by the shadow. After propagating these early freeze-out
partons to $t=5$ fm/c at 200 GeV and $t=4$ fm/c at 39 GeV, the
density contour in the reaction plane (x-o-z) plane is shown in
Fig.~\ref{F12}. We found that partons at $x\cdot z>0$ ($x\cdot z<0
$) contribute positively (negatively) to the directed flow. Although
there are more partons at $x\cdot z>0$ ($x\cdot z<0 $) for 200 GeV
(39 GeV), which leads to the net positive (negative) directed flow
slope near midrapidity, the local density is slightly higher at
$x\cdot z<0$ ($x\cdot z>0 $) for 200 GeV (39 GeV), leading to the
change of the $v_1$ slope sign from a pure coordinate coalescence,
especially for baryons formed by three quarks close in coordinate
space.

\subsection{Effect of hadronic evolution on $v_1$}

\begin{figure}[h]
\includegraphics[scale=0.38]{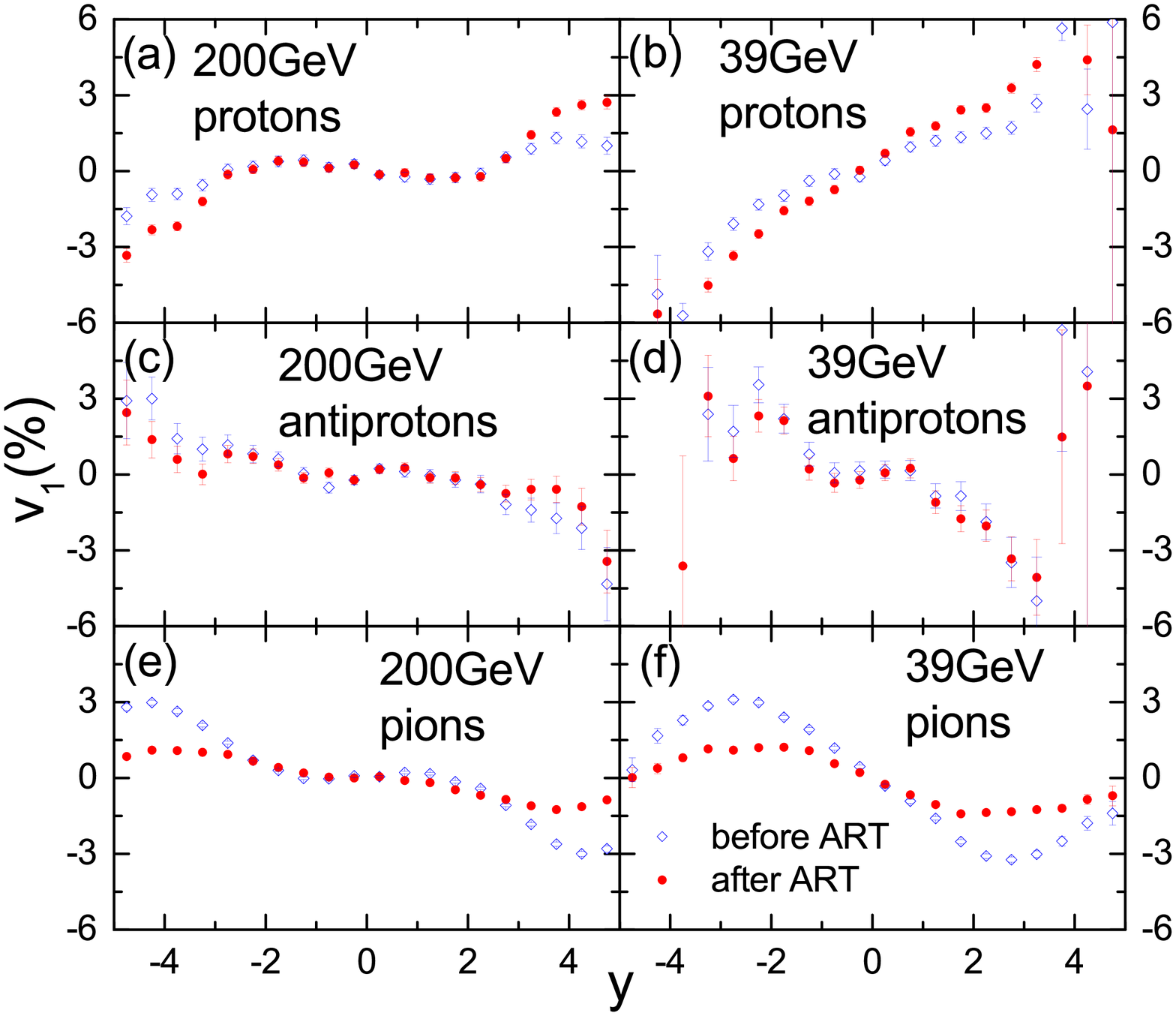}
\caption{(Color online) Directed flow ($v_1$) of protons (top), antiprotons (middle), and charged
pions (bottom) versus rapidity ($y$) before and after hadronic rescatterings
in midcentral Au+Au collisions at $\sqrt{s_{NN}}$ = 200 GeV (left) and 39 GeV (right).} \label{F10}
\end{figure}

The directed flow of hadrons after hadronzation presented in the
previous subsection is further modified by the hadronic evolution
described by ART, containing various elastic, inelastic, and decay
channels. To illustrate the effect of hadronic rescatterings on
$v_1$ in the AMPT model, we present in Fig.~\ref{F10} the directed
flow of initial (before ART) and final (after ART) protons (top panels), antiprotons (middle panels), and
charged pions (bottom panels) in midcentral Au+Au collisions at $\sqrt{s_{NN}}$ =
200 GeV and 39 GeV. After hadronic rescatterings, it is seen that
the slope of the directed flow generally becomes less negative or
increases, while the sign of the $v_1$ slope near midrapidity region
is mostly not changed. The effect of the hadronic evolution on $v_1$ is
seen to be larger at lower collision energies compared to that at
higher collision energies.

\section{\label{sec:4} SUMMARY AND OUTLOOK}

Based on the framework of a multiphase transport model, we have
discussed some interesting topics relevant to how the directed flow
is developed in relativistic heavy-ion collisions, which have not
been well addressed previously. As the partonic phase evolves, a
non-monotonic behavior of the directed flow is observed at higher
collision energies, and the later dynamics is able to change the
slope sign of the directed flow at midrapidity region, due to the
transfer of partons among different rapidity regions, but this is
not observed at lower collision energies as a result of shorter life
time of the partonic phase. The splitting of the directed flow for
various quark species is observed at lower collision energies due to
their different initial rapidity-odd eccentricities, while such
splitting becomes very small at higher collision energies. The
directed flow of hadrons is a result of competition between the
coalescence in momentum and coordinate space, with the weight
determined by the Gaussian width of the Wigner function in the
dynamical coalescence scenario, and is further modified by the
hadronic rescatterings. The coalescence mechanism as well as the hadronic rescatterings discussed in the present manuscript can be possible reasons accounting for the violation of the number of constituent quark scaling for the directed flow mentioned in Ref.~\cite{STAR17}.

In the future study, we will improve the coalescence and introduce
the mean-field potentials, or similarly, the equation of
state~\cite{hydro,transport,antiflow,thirdflow}, to the multiphase
transport model as in Refs.~\cite{MF,Xu16}. It will be of great
interest to study quantitatively how the results obtained in this
work are modified by the mean-field potentials, which are expected
to be different for different quark and hadron species. Based on the
studies of the directed flow from particle scatterings in the
present study as well as the mean-field potentials in the future
study, one should be able have a better understanding of the
directed flow at various collision energies and for various particle
species, thus hopefully extract useful information of the
hadron-quark phase transition and the QCD phase diagram.

\begin{acknowledgements}
We thank Chen Zhong for maintaining the high-quality performance of
the computer facility. This work was supported by the Major State
Basic Research Development Program (973 Program) of China under
Contract Nos. 2015CB856904 and 2014CB845401, the National Natural
Science Foundation of China under Grant Nos. 11475243 and 11421505,
the ``100-talent plan" of Shanghai Institute of Applied Physics under
Grant Nos. Y290061011 and Y526011011 from the Chinese Academy of
Sciences, and the Shanghai Key Laboratory of Particle Physics and
Cosmology under Grant No. 15DZ2272100.
\end{acknowledgements}

\appendices

\renewcommand\thesection{APPENDIX~\Alph{section}}
\renewcommand\theequation{\Alph{section}.\arabic{equation}}
\section{Dynamical coalescence for hadrons with Wigner function}
\label{app}

In the dynamical coalescence model, the probability for a pair of
quark and antiquark to form a meson is proportional to the quark
Wigner function of the meson times the statistical factor, i.e.,
\begin{eqnarray}\label{fm}
f_M(\bm{\rho},\bm{k}_{\rho}) &=&
8g_M\exp\left(-\frac{\bm{\rho}^2}{\sigma_{\rho}^2}-\bm{k}_{\rho}^2\sigma_{\rho}^2\right),
\end{eqnarray}
where $g_M= 1/36$ is the statistical factor for pions, and
\begin{eqnarray}
\bm{\rho} &=& \frac{1}{\sqrt{2}}(\bm{r}_1-\bm{r}_2),\\
\bm{k}_{\rho} &=& \sqrt{2}\frac{m_2\bm{k}_1-m_1\bm{k}_2}{m_1+m_2}
\end{eqnarray}
are the relative distance in the coordinate and momentum space for
the two-particle system, with $m_i$, $\textbf{r}_i$, and
$\textbf{k}_i$ being the mass, coordinate, and momentum of the $i$th
particle, respectively. The width parameter $\sigma_{\rho}$ is
related to the root-mean-square (RMS) radius of the meson through
the relation
\begin{eqnarray}
\langle r_M^2 \rangle &=& \frac{3}{2}\frac{m_1^2+m_2^2}{(m_1+m_2)^2}\sigma_{\rho}^2\notag\\
&=& \frac{3}{4}\frac{m_1^2+m_2^2}{m_1m_2(m_1+m_2)\omega},
\end{eqnarray}
where the second line follows if we use the relation
$\sigma_\rho=1/\sqrt{\mu_1\omega}$ in terms of the oscillator
frequency $\omega$ and the reduced mass $\mu_1=2(1/m_1+1/m_2)^{-1}$.

Similarly, the probability for three light quarks to form a baryon
is expressed as
\begin{eqnarray}\label{fb}
\begin{split}
&f_B(\bm{\rho},\bm{\lambda},\bm{k}_{\rho},\bm{k}_{\lambda}) \\&=
8^2g_B\exp\left(-\frac{\bm{\rho}^2}{\sigma_{\rho}^2}-\frac{\bm{\lambda}^2}{\sigma_{\lambda}^2}-\bm{k}_{\rho}^2\sigma_{\rho}^2-\bm{k}_{\lambda}^2\sigma_{\lambda}^2\right),
\end{split}
\end{eqnarray}
where $g_B=1/108$ is the statistical factor for protons, and
\begin{eqnarray}
\bm{\lambda} &=& \sqrt{\frac{2}{3}}\left(\frac{m_1\bm{r}_1+m_2\bm{r}_2}{m_1+m_2}-\bm{r}_3\right),\\
\bm{k}_{\lambda} &=& \sqrt{\frac{3}{2}}\frac{m_3(\bm{k}_1+\bm{k}_2)-(m_1+m_2)\bm{k}_3}{m_1+m_2+m_3}
\end{eqnarray}
are the relative distance in the coordinate and momentum space
between the third particle and the system formed by the first and
the second particles. The width parameter $\sigma_{\lambda}$ is
related to the oscillator frequency via $1/\sqrt{\mu_2\omega}$, with
$\mu_2=(3/2)[1/(m_1+m_2)+1/m_3]^{-1}$. The RMS radius of the baryon
is then given by
\begin{eqnarray}
\begin{split}
&\langle r_B^2 \rangle = \\
&\frac{1}{2}\frac{m_1^2(m_2+m_3)+m_2^2(m_1+m_3)+m_3^2(m_1+m_2)}{(m_1+m_2+m_3)m_1m_2m_3\omega}.
\end{split}
\end{eqnarray}
The RMS radius of the produced hadron is taken from Ref.~\cite{RMS},
which is 0.61 fm for $\pi^+$ and 0.877 fm for protons, respectively.

\renewcommand\thesection{APPENDIX~\Alph{section}}
\renewcommand\theequation{\Alph{section}.\arabic{equation}}
\section{Naive coalescence scenario}
\label{naive}

In the naive coalescence scenario, the momentum distribution of quarks
inside hadrons is neglected. A meson with momentum $\bm{p}$ are
formed by a pair of quark and antiquark with half the meson
momentum $\bm{p}/2$ co-moving in the same direction. This is
actually the limit of the dynamical coalescence with $\omega=0$ or
infinitely large Gaussian width in the Wigner function. In the
following, we briefly remind the relation between collective flows
of hadrons and their constituent quarks in the naive coalescence
scenario as in Refs.~\cite{naive1,naive2}.

The azimuthal distribution of mesons in momentum space can be expressed as
\begin{eqnarray}\label{ffm}
F(\phi, \bm{p})  \propto 1+2\sum\limits_{n=1}^{\infty}V_n(\bm{p})\cos(n\phi) \propto f(\phi, \bm{p}/2)^2,
\end{eqnarray}
where $f(\phi)$ is azimuthal distribution function of partons, i.e.,
\begin{eqnarray}\label{ffq}
f(\phi, \bm{p}/2) \propto 1+2\sum\limits_{n=1}^{\infty}v_n(\bm{p}/2)\cos(n\phi).
\end{eqnarray}
From Eqs.~(\ref{ffm}) and (\ref{ffq}), the $n$th-order anisotropy
flow $V_n(\bm{p})$ of mesons can be expressed in terms of the parton
anisotropy flow $v_n(\bm{p}/2)$ as
\begin{eqnarray}
V_n=\frac{1}{N}(2v_n+\sum\limits_{i=1}^{n-1}v_iv_{n-i}+2\sum\limits_{i=1}^{\infty}v_iv_{n+i}),
\end{eqnarray}
with $N=1+2\sum\limits_{i=1}^{\infty}v_i^2$.

Similarly, the azimuthal distribution for baryons in momentum space
can be expressed as the third power of the azimuthal distribution
for partons, i.e.,
\begin{eqnarray}
\tilde{F}(\phi, \bm{p}) \propto 1+2\sum\limits_{n=1}^{\infty}\tilde{V}_n(\bm{p})\cos(n\phi) \propto f(\phi, \bm{p}/3)^3.
\end{eqnarray}
The $n$th-order anisotropy flow $\tilde{V}_n$ of baryons is
expressed in terms of the parton anisotropy flow $v_n(\bm{p}/3)$ as
\begin{eqnarray}
\tilde{V}_n&=&\frac{1}{\tilde{N}}(3v_n+3\sum\limits_{i=1}^{n-1}v_iv_{n-i}+6\sum\limits_{i=1}^{\infty}v_iv_{n+i}\notag\\
&+&3\sum\limits_{i=1}^{\infty}\sum\limits_{j=1}^{\infty}v_iv_jv_{n+i+j}+3\sum\limits_{i=1}^{\infty}\sum\limits_{j=1}^{n+i-1}v_iv_jv_{n+i-j}\notag\\
&+&\sum\limits_{i=1}^{n-1}\sum\limits_{j=1}^{n-i-1}v_iv_jv_{n-(i+j)}),
\end{eqnarray}
with $\tilde{N}=1+6\sum\limits_{i=1}^{\infty}v_i^2+6\sum\limits_{i=1}^{\infty}\sum\limits_{j=1}^{\infty}v_iv_jv_{i+j}$.

Neglecting the higher-order terms, the scaling relations between the
directed flows of baryons ($\tilde{V}_1$), mesons ($V_1$), and
partons ($v_1$) are
\begin{eqnarray}
V_1(\bm{p}) &\approx& \frac{2v_1(\bm{p}/2)}{1+2v_1^2(\bm{p}/2)} \approx 2v_1(\bm{p}/2),\\
\tilde{V}_1(\bm{p}) &\approx& \frac{3v_1(\bm{p}/3)}{1+6v_1^2(\bm{p}/3)} \approx 3v_1(\bm{p}/3).\\\notag
\end{eqnarray}
Note that the scaling relation for the rapidity-even directed flow
or the higher-order anisotropic flows ($n>2$) is often discussed at
midrapidities ($y \approx 0$), where the momentum
$\bm{p}$ can be approximated by the transverse momentum $p_T$. In the most
general case, the momentum $\bm{p}$ is a vector related to both the
rapidity $y$ and the transverse momentum $p_T$, as in the present
study of the rapidity-odd directed flow.

\end{CJK*}

\begin{thebibliography}{99}
\bibitem{Ars05} I. Arsene {\it et al.} (PHOBOS Collaboration), Nucl. Phys.
\textbf{A757}, 1 (2005).

\bibitem{Bac05} B. B. Back {\it et al.} (BRAHMS Collaboration), Nucl. Phys.
\textbf{A757}, 28 (2005).

\bibitem{Ada05a} J. Adams {\it et al.} (STAR Collaboration), Nucl. Phys.
\textbf{A757}, 102 (2005).

\bibitem{Adc05} K. Adcox {\it et al.} (PHENIX Collaboration), Nucl. Phys.
\textbf{A757}, 184 (2005).

\bibitem{v11} S. Singha {\it et al.}, Advances in High Energy Physics {\bf 16}, 2836989 (2016).

\bibitem{voe1} D. Teaney and L. Yan, Phys. Rev. C {\bf 83}, 064904 (2011).

\bibitem{voe2} M. Luzum and J. Y. Ollitrault, Phys. Rev. Lett. {\bf 106}, 102301 (2011).

\bibitem{RHIC} L. Adamczyk {\it et al.} (STAR Collaboration), Phys. Rev. Lett. {\bf 112}, 162301 (2014).

\bibitem{Ste14} J. Steinheimer {\it et al.}, Phys. Rev. C {\bf 89}, 054913 (2014).

\bibitem{Kon14} V. P. Konchakovski {\it et al.}, Phys. Rev. C {\bf 90}, 014903 (2014).

\bibitem{Nar16} Y. Nara {\it et al.}, Phys. Rev. C {\bf 94}, 034906 (2016).

\bibitem{Nar17} Y. Nara, H. Niemi, J. Steinheimer, and H. St\"ocker, Phys. Lett. B {\bf 769}, 543 (2017).

\bibitem{Iva15} Yu. B. Ivanov and A. A. Soldatov, Phys. Rev. C \textbf{91}, 024915 (2015).

\bibitem{Bat16} P. Batyuk, D. Blaschke, M. Bleicher, Yu. B. Ivanov, Iu. Karpenko, S. Merts, M. Nahrgang, H. Petersen, and O. Rogachevsky, Phys. Rev. C \textbf{94}, 044917 (2016).

\bibitem{AMPT} Z. W. Lin, C. M. Ko, B. A. Li, B. Zhang, and S. Pal, Phys. Rev. C {\bf 72}, 064901 (2005).

\bibitem{chenjy} J. Y. Chen, J. X. Zuo, X. Z. Cai, F. Liu, Y. G. Ma, and A. H. Tang, Phys. Rev. C {\bf 81}, 014904 (2010).

\bibitem{hijing} X. N. Wang and M. Gyulassy, Phys. Rev. D {\bf 44}, 3501 (1991).

\bibitem{zpc} B. Zhang, Comput. Phys. Commun. {\bf 109}, 193 (1998).

\bibitem{art} B. A. Li and C. M. Ko, Phys. Rev. C {\bf 52}, 2037 (1995).

\bibitem{dynamic} V. Greco, C. M. Ko, and P. Levai, Phys. Rev. Lett. {\bf 90}, 202302 (2003); Phys. Rev. C {\bf 68}, 034904 (2003).

\bibitem{Liu99} H. Liu, S. Panitkin, and N. Xu, Phys. Rev. C {\bf 59}, 348 (1999).

\bibitem{PHOBOSv1} B. B. Back {\it et al.} (PHOBOS Collaboration), Phys. Rev. Lett. {\bf 97}, 012301 (2006).

\bibitem{STARv1} B. I. Abelev {\it et al.} (STAR Collaboration), Phys. Rev. Lett. {\bf 101}, 252301 (2008).

\bibitem{RHICv2} L. Adamczyk {\it et al.} (STAR Collaboration), Phys. Rev. Lett. \textbf{110}, 142301 (2013).

\bibitem{Dun11} J. C. Dunlop, M. A. Lisa, and P. Sorensen, Phys. Rev. C \textbf{84}, 044914 (2011).

\bibitem{Guo12} Y. Guo, F. Liu, and A. H. Tang, Phys. Rev. C \textbf{86}, 044901 (2012).

\bibitem{epsilon1} B. Alver and G. Roland, Phys. Rev. C {\bf 81}, 054905 (2010).

\bibitem{epsilon2} R. A. Lacey {\it et al.}, Phys. Rev. C {\bf 83}, 044902 (2011).

\bibitem{epsilon3} H. Petersen, G. Y. Qin, S. A. Bass, and B. M\"{u}ller, Phys. Rev. C {\bf 82}, 041901(R) (2010).

\bibitem{wiggle} R. J. M. Snellings, H. Sorge, S. A. Voloshin, F. Q. Wang, and N. Xu, Phys. Rev. Lett. {\bf 84}, 2803 (2000).

\bibitem{splitting} C. M. Ko, T. Song, F. Li, V. Greco, and S. Plumari, Nucl. Phys. A {\bf 928}, 234 (2014).

\bibitem{naive1} P. F. Kolb {\it et al.}, Phys. Rev. C {\bf 69}, 051901(R) (2004).


\bibitem{naive2} J. Y. Jia and C. Zhang, Phys. Rev. C {\bf 75}, 031901(R) (2007).

\bibitem{STAR17} L. Adamczyk {\it et al.}, [STAR Collaboration], arXiv: 1708.07132 [hep-ex].

\bibitem{hydro} H. St\"{o}cker, Nucl. Phys. A {\bf 750}, 121 (2005).

\bibitem{transport} S. A. Bass {\it et al.}, Prog. Part. Nucl. Phys. {\bf 41}, 255 (1998).

\bibitem{antiflow} J. Brachmann {\it et al.}, Phys. Rev. C {\bf 61}, 024909 (2000).

\bibitem{thirdflow} L. P. Csernai and D. R\"{o}hrich, Phys. Lett. B {\bf 458}, 454 (1999).

\bibitem{MF} J. Xu, T. Song, C. M. Ko, and F. Li, Phys. Rev. Lett. {\bf 112}, 012301 (2014).

\bibitem{Xu16} J. Xu and C. M. Ko, Phys. Rev. C {\bf 94}, 054909 (2016).

\bibitem{RMS} J. Beringer {\it et al.} [Particle Data Group Collaboration], Phys. Rev. D {\bf 86}, 010001 (2012).
\end{thebibliography}
\end{document}